\documentclass[twocolumn,trackchanges]{aastex701}
\usepackage[utf8]{inputenc}
\usepackage{xcolor}

\usepackage{booktabs} %

\newcommand{\teff}{\mbox{$T_{\rm eff}$}}

\newcommand{\feh}{\mbox{$\rm{[Fe/H]}$}}
\newcommand{\lum}{\mbox{$\rm{L}$}}

\begin{document}

\title{Insights into Planet Formation from the Ages, Masses, and Elemental Abundances of Host Stars}

\author[orcid=0000-0003-3957-9067]{Xunzhou Chen}
\affiliation{School of Science, Hangzhou Dianzi University, Hangzhou, PR China}
\affiliation{National Astronomical Data Center Zhijiang Branch, Hangzhou, PR China}
\email{cxz@hdu.edu.cn}

\author[orcid=0000-0003-0795-4854]{Tiancheng Sun}
\affiliation{CAS Key Laboratory of Optical Astronomy, National Astronomical Observatories, Chinese Academy of Sciences, Beijing 100101, China}
\email[show]{suntc@bao.ac.cn}

\author[orcid=0009-0009-1338-1045]{Lifei Ye}
\affiliation{School of Physics and Astronomy, Beijing Normal University, Beijing, China}
\email{201831160010@mail.bnu.edu.cn}

\begin{abstract}
How planetary systems form and evolve is a key question in astronomy. Revealing how host star properties—such as elemental abundances, age, and mass—differ from those of non-host stars, and how they correlate with planetary characteristics such as radius, provides new insights into the formation and evolutionary pathways of planetary systems. We determine precise ages for 18890 dwarfs and subgiants from the LAMOST--Kepler--Gaia sample with mean age uncertainty $\sim$15\% (median $\sim$10\%). Within the framework of Galactic chemical evolution, we find that $\sim$86\% of planet-hosting stars younger than 8 Gyr occupy the upper branch ([Fe/H] $> -0.2$) of the characteristic ``V-shape" age–metallicity relation of the Galactic disk. Based on guiding radii ($R_{\rm g}$), we further infer that $\sim$19\% of these young hosts likely originated in the inner disk and subsequently migrated to the solar neighborhood. Among stars older than 10 Gyr, host stars tend to be more metal-rich, with nearly 59\% having [Fe/H] $>-0.2$. This suggests that both young and old planet-hosting stars preferentially form in relatively metal-rich environments. However, for host stars with [Fe/H] $< -0.2$, we find that their [Fe/H] are on average lower by $\sim$0.16\,dex compared to non-host stars of similar age and mass, indicating that [Fe/H] is unlikely to be the dominant factor governing planet formation in metal-poor environments. We also identify systematic depletion of volatile elements—especially carbon—in planet hosts. Moreover, host star [Fe/H] exhibits a weak correlation with planet radius, while [$\alpha$/Fe] primarily supports the formation of small planets.


\end{abstract}


\section{Introduction} 
Understanding when, where, and how planets form remains a central question in exoplanet science. Since the discovery of the first exoplanet orbiting a main-sequence star—51~Pegasi~b—by \citet{1995Natur.378..355M}, the field has grown rapidly. Both ground-based programs \citep[e.g.,][]{2011arXiv1109.2497M} and space-based missions \citep[e.g.,][]{2010Sci...327..977B} have contributed to a catalog of over 5000 confirmed exoplanets (NASA Exoplanet Archive; \citealt{2013PASP..125..989A}), enabling statistically robust studies of planetary systems and their host stars.

A fundamental goal is to understand how the stellar properties—mass, metallicity, and age—differ between stars with and without detected planets. Spectroscopic surveys \citep[e.g.,][]{2012MNRAS.423..122B,2013ApJ...771..107E,2014Natur.509..593B,2014ApJ...789L...3D,2015AJ....149..143F,2018AJ....155...89P,2021AJ....161..114S,2022AJ....164...60S} have established that metallicity ([Fe/H]) correlates with planet occurrence, particularly for giant planets, which tend to orbit younger and more metal-rich stars. These trends are often attributed to either primordial disk conditions that favor planet formation in metal-rich environments \citep[e.g.,][]{2004A&A...415.1153S,Johnson_2010}, or to pollution via the accretion of planetary material \citep[e.g.,][]{1996Natur.380..606L,1997ApJ...491L..51L}.

Among space-based surveys, Kepler stands out for its long-term, high-precision monitoring of nearly 200000 stars \citep{2017ApJS..229...30M,2020AJ....159..280B}, leading to the discovery of more than 2000 confirmed planets and thousands of candidates \citep{Thompson_2018}. Its broad spatial coverage—spanning several kiloparsecs and diverse Galactic environments—makes it uniquely suited for studying planets and their hosts in a Galactic context \citep[e.g.,][]{2021ApJ...909..115C,2021AJ....162..100C,2023AJ....166..243Y}. Such studies require not only precise planet detections, but also accurate characterization of the full stellar sample—particularly the stellar ages, which are key to understanding the temporal dimension of planet formation and evolution. However, stellar age determination remains challenging. Most Kepler stars have well-measured parameters—mass to $\sim$7\%, radius to $\sim$4\%, and effective temperature within 112~K—while stellar ages remain uncertain, with a mean error $\sim$56\% from isochrone fitting \citep{2020AJ....159..280B}.

Recent advances in artificial intelligence have enabled the derivation of high-precision stellar atmospheric parameters from low-resolution spectra. Combined with the exquisite astrometric data from Gaia DR3 \citep{2023A&A...674A...1G}, these developments now allow stellar ages to be estimated with unprecedented accuracy. Using the data-driven Payne method \citep[DD-Payne;][]{2019ApJS..245...34X} applied to LAMOST spectra and trained on stars in common with APOGEE \citep{2017AJ....154...94M} and GALAH \citep{2015MNRAS.449.2604D}, \citet{2022Natur.603..599X} determined atmospheric parameters for nearly 250000 subgiants and derived stellar ages with a median precision of 7.5\%.
Similarly, incorporating Gaia luminosities, \citet{2023MNRAS.523.1199S,2023ApJS..268...29S} determined ages for $\sim$40000 main-sequence turnoff stars from GALAH DR3 atmospheric parameters \citep{2021MNRAS.506..150B} (median uncertainty $\sim$9.4\%) and for $\sim$67000 dwarfs from LAMOST DR5 atmospheric parameters \citep{2019ApJS..245...34X} (median uncertainty $\sim$16\%). These advances demonstrate that precise stellar parameters now allow for more detailed investigations of correlations between planets and stars.

In this study, we investigate the correlations between host stars and their planets in the LAMOST–Kepler–Gaia field. Using stellar evolutionary models, we determine precise ages and masses for a sample of nearly 19000 dwarfs and subgiants, including 392 planetary systems hosting a total of 534 planets. We then compare the distributions of stellar parameters between host and non-host stars and examine the relationships between stellar properties and planetary radii. The paper is organized as follows. Section \ref{sec:style} describes the sample selection and age determination in detail. Section \ref{sec:floats} presents our results. Section \ref{sec:summary} summarizes our main conclusions.

\section{Data} \label{sec:style}
\subsection{Sample Selection}
We utilize stellar atmospheric parameters and elemental abundances from the LAMOST DR9 DD-Payne catalogue\footnote{https://zenodo.org/records/15254859} \citep{2025ApJS..279....5Z}, which provides homogeneous measurements for FGK-type stars, including 22 elemental abundances. The [Fe/H] values are corrected for non-local thermodynamic equilibrium (non-LTE) effects, effective temperatures are calibrated to the infrared flux method (IRFM) scale, and surface gravities are validated against asteroseismic constraints. We adopt parallaxes, radial velocities, and proper motions from Gaia DR3 \citep{2023A&A...674A...1G}, which provide high-precision astrometry and kinematics, and planetary properties from the Kepler DR25 catalog \citep{Thompson_2018, k2pandc}. The Kepler Stellar Properties Catalog \citep{2020AJ....159..280B} is also used for source selection. 

We start with 186301 stars from \citet{2020AJ....159..280B}, which are then cross-matched to the LAMOST DR9 DD-Payne catalog within $1.5^{\prime\prime}$, yielding 76058 stars. These stars are subsequently matched to Gaia DR3, resulting in a final sample of 74581 stars with combined Kepler, LAMOST, and Gaia data. We apply quality cuts to ensure reliable stellar and planetary data: signal-to-noise ratio in the G band (SNRG $\geq 20$, 60971 stars) and $\chi^2$ flag ($\leq 3$, 54057 stars) from DR9 DD-Payne to select precise and reliable stellar parameters; Gaia RUWE $\leq 1.2$ (44396 stars) to exclude binaries; valid abundance flags for C, O, Na, Mg, Si, Ca, and Ti (44347 stars) from DD-Payne to retain reliable elemental abundances; and available Combined Differential Photometric Precision (CDPP) (44148 stars) from Kepler DR25 to ensure transit search completeness. We then select stars with $T_{\rm eff}=5000$--$6800$\,K, as the upper limit follows \citet{2022Natur.603..599X} and the lower limit corresponds to the temperature range of most dwarf calibration stars in \citet{2025ApJS..279....5Z}, yielding 33219 stars.

To robustly infer stellar ages for our sample, precise determinations of stellar luminosities are required. We cross-match the sample with the Two Micron All Sky Survey (2MASS) \citep{2003yCat.7233....0S,2006AJ....131.1163S} to obtain K-band apparent magnitudes, and apply parallax zero-point corrections to Gaia DR3 using the \texttt{gaiadr3-zeropoint} code described by \cite{2021A&A...649A...4L}. By combining the 2MASS K-band magnitudes, the corrected Gaia DR3 parallaxes, and extinction values from the \cite{2019ApJ...887...93G} dust map, we derive luminosities of 33212 stars using the direct method implemented in the isoclassify code \citep{2017ApJ...844..102H,2020AJ....159..280B,2023arXiv230111338B}.

Guided by the goal of understanding how a host star’s birthplace influences planet formation, we estimate the guiding radius ($R_{\rm guiding}$), which corresponds to the average orbital radius around the Galactic center. Unlike the instantaneous Galactocentric distance, $R_{\rm guiding}$ averages over radial oscillations along the orbit, a process known as ``blurring", and therefore provides a more reliable proxy for a star’s likely birth radius. Stars near the Sun may have diverse origins, having migrated from the inner or outer disk or formed locally. By examining $R_{\rm guiding}$, we can trace the probable birth locations and explore potential correlations between birth environment of host stars and planet formation.

We compute $R_{\rm guiding}$ using the geometric distances from \citet{2021AJ....161..147B}, as well as radial velocities and proper motions from Gaia DR3. The calculations are performed with the \texttt{Galpy} \citep{2015ApJS..216...29B}, adopting the \texttt{MWPotential2014} potential. We assume the Sun is located at $R_{\odot} = 8.21$ kpc \citep{2017MNRAS.465...76M} and $Z_{\odot} = 25$ pc \citep{2008ApJ...673..864J} above the Galactic midplane, with a local standard of rest (LSR) velocity of 248.27 km s$^{-1}$ \citep{2004ApJ...616..872R}. The solar motion relative to the LSR is taken as $(U_{\odot}, V_{\odot}, W_{\odot}) = (11.1, 15.17, 7.25)$ km s$^{-1}$ \citep{2010MNRAS.403.1829S}, where the adopted value of $V_{\odot}$ is consistent with that used by \citet{2021MNRAS.506..150B}.

A small subset of the sample ($\sim$12\%) lacks Gaia radial velocity measurements. For these stars, we cross-match with the LAMOST DR9 catalog \footnote{https://www.lamost.org/dr9/v2.0/} to obtain radial velocities. To correct for the known systematic offset of 5.38 km s$^{-1}$ \citep{2015ApJ...809..145T,2017MNRAS.472.3979S,2019ApJ...871..184T} between LAMOST and Gaia measurements, we apply an additive correction of +5.38 km s$^{-1}$ to the LAMOST values.

\begin{figure}[htbp]
  \centering
  \includegraphics[width=0.45\textwidth]{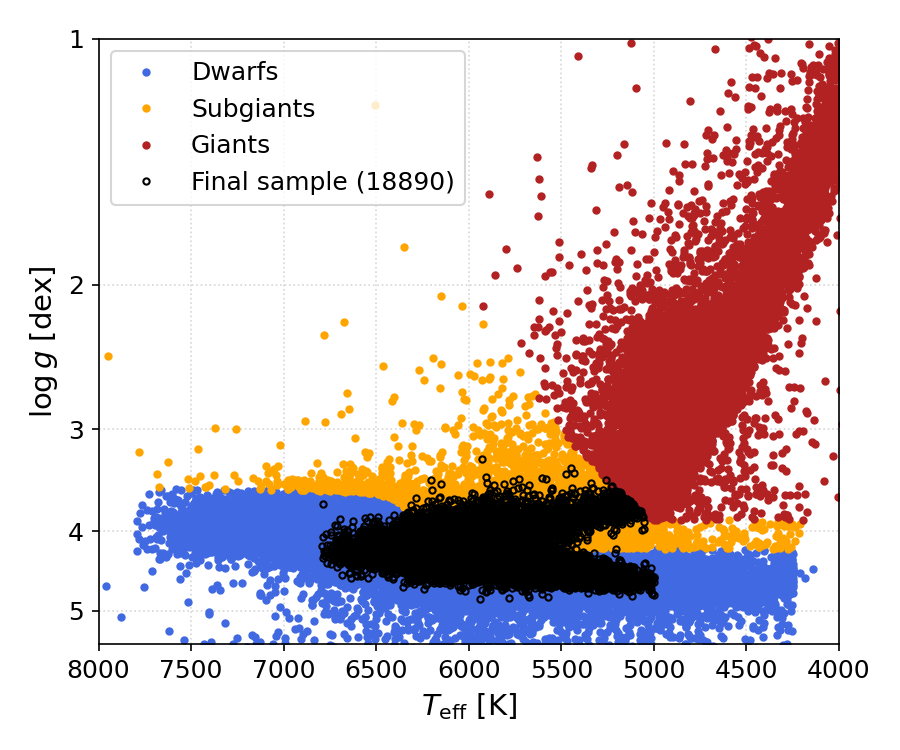}
  \caption{
  Kiel diagram of all LAMOST-Kepler-Gaia sample stars. Dwarfs, subgiants, and giants are shown in blue, orange, and red, respectively, based on classification with the \texttt{evolstate} algorithm. The black open circles indicate the location of our final sample of 18890 stars with well-constrained ages.
  }
  \label{fig:hr}
\end{figure}

Following \citet{2020AJ....159..279B}, we use the \texttt{evolstate} Python package to identify stellar evolutionary stages and exclude giants, retaining 31502 dwarfs and subgiants. Although previous studies have suggested that subgiants may experience surface abundance alterations due to mixing processes, potentially introducing systematic biases in chemical analyses \citep[e.g.,][]{2022AJ....164...60S}, most subgiants have shallow convective envelopes that do not experience dredge-up, thereby preserving surface abundances comparable to those on the main sequence. Additionally, \citet{2020A&A...633A..23D} showed that atomic diffusion significantly alters [Fe/H] and [C/H] in stars more massive than 1.44~$M_\odot$. To avoid these effects, we restrict our sample to stars with $M < 1.4~M_\odot$, using stellar mass estimated by \citet{2020AJ....159..280B} for initial selection, resulting in a sample of 27676 stars.

\begin{figure*}[htbp]
  \centering
  \includegraphics[width=\textwidth]{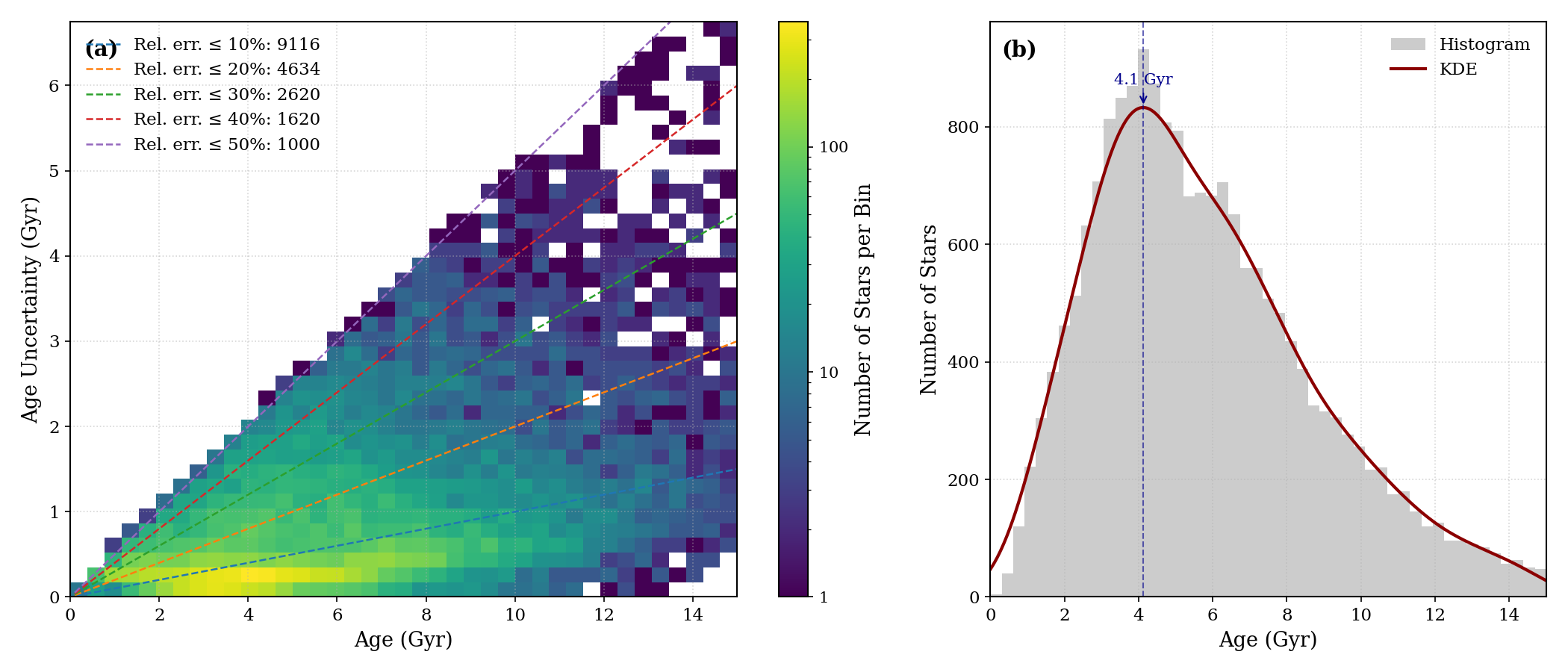}
  \caption{
  Age precision and distribution of 18990 sample stars. 
  \textbf{(a)} Two-dimensional histogram of stellar age versus age uncertainty, with dashed lines indicating fractional age uncertainties of 10\%, 20\%, 30\%, 40\%, and 50\%. 
  The colorbar indicates the number of stars per bin in logarithmic scale. 
  \textbf{(b)} Age histogram (gray bars) overlaid with a KDE-smoothed curve (red). Vertical dashed lines and arrows highlight the detected peaks in the age distribution.
  }
  \label{fig:age_distribution}
\end{figure*}
\subsection{Fundamental Parameter Estimation}

In order to obtain fundamental parameters including stellar age, we use a Bayesian scheme which is similar as \citet{2010A&A...522A...1K}, and \citet{2010ApJ...710.1596B} to find the most probable stellar models from evolutionary tracks. Based on a set of observed constrains $\o$ (in our case, they are \teff, \lum, and \feh), we define the likelihood that matches the observed constrains as:
\begin{equation}\label{eq4}
L=\frac{1}{\sqrt{2\pi}\sigma}exp{(\frac{-\chi^2}{2})},
\end{equation}
where
\begin{equation}\label{eq5}
\chi^2=(\frac{\o_{obs}-\o_{model}}{\sigma})^2.
\end{equation}
Here $\sigma$ is the error of the observation $\o_{obs}$.
According to Bayes' theorem, the posterior probability of model $M_{i}$ given data $D$ is computed via:
\begin{equation}\label{eq6}
p(M_{i}|D,I)=\frac{p(M_{i}|I)p(D|M_{i},I)}{p(D|I)}.
\end{equation}
We assume a uniform prior $p(M_{i}|I)\ =\ \frac{1}{N_m}$, where $N_m$ is the total number of computed models. Our likelihood function is defined as:
\begin{equation}\label{eq7}
p(D|M_{i},I)=L(\teff,\lum, \feh)=L_{\teff}L_{\lum}L_{\feh}.
\end{equation}
Since $p(D|I)$ is just a normalization factor and $p(M_{i}|I)$ is constant, we have:
\begin{equation}\label{eq8}
p(M_{i}|D,I) \propto p(D|M_{i},I).
\end{equation}

We then fit a Gaussian function to the likelihood distribution, adopting the mean as the estimate and the standard deviation as its uncertainty.

Based on the above sample of 27676 stars, we derived fundamental stellar parameters by matching the observed values of \teff, \lum, and [Fe/H] to the $\alpha$-enhanced stellar evolution models developed by \citet{2023MNRAS.523.1199S}. Including $\alpha$-enhancement in stellar evolution models is essential because $\alpha$-elements substantially alter stellar opacities, thereby influencing the internal structure and evolutionary pathways of stars. Stellar tracks and isochrones computed with $\alpha$-enhanced compositions exhibit systematically hotter and bluer turn-offs in the Hertzsprung–Russell diagram compared with their scaled-solar counterparts. These differences have a direct impact on age estimates for high-$\alpha$ stars, with $\alpha$-enhanced models typically yielding younger ages than those inferred under scaled-solar assumptions. These systematic effects have been demonstrated in multiple studies \citep{2000ApJ...532..430V,2001ApJ...556..322B,2001ApJS..136..417Y,2002ApJS..143..499K,2012ApJ...755...15V}, underscoring that $\alpha$-enhancement must be accounted for when modeling stellar populations with high $\alpha$-abundance. Given that our sample includes both low- and high-$\alpha$ stars, the adoption of $\alpha$-enhanced evolution models in our analysis is critical for obtaining physically robust and self-consistent results.

For each star, we employed a subset of models corresponding to the closest [$\alpha$/Fe] value, thereby ensuring consistency with the observed level of $\alpha$-enhancement. The [$\alpha$/Fe] ratio for each star was calculated as the error-weighted mean of the individual abundances of the $\alpha$-elements (Mg, Si, Ca, and Ti) from the DR9 DD-Payne catalog.

Applying selection criteria including a relative age uncertainty below 50\%, an absolute age less than 15~Gyr, an age $-$ 2*age$_{\rm error}$ below the age of the Universe ($<13.8$~Gyr\footnote{Planck Collaboration XIII \citep{2016A&A...594A..13P}}), and stellar mass below 1.4~$M_\odot$, we obtained a final sample of 18990 stars.
To assess the accuracy of the age determinations in our sample, we cross-matched the 18990 stars with the subgiant sample from \citet{2022Natur.603..599X}, resulting in 2866 common sources, and present a comparison of stellar ages between this work and the results of \citet{2022Natur.603..599X} (see Appendix~\ref{app:age_comparison}).

\subsection{Planet Sample}

Our Kepler planet sample is based on the Kepler DR25 catalog \citep{Thompson_2018, k2pandc}. We select planet candidates associated with stars in our stellar sample and exclude those flagged as false positives. We limit the analysis to planets with orbital periods shorter than 100 days, as both the number of detected planets and the detection efficiency decline significantly beyond this threshold \citep{2017ksci.rept...19B}. Additionally, planets with radii smaller than 0.5~$R_\oplus$ are excluded due to their low detection efficiency \citep{2017ksci.rept...19B}.

After applying these criteria, we obtain a final sample of 18890 stars with precise age estimates, including 392 planetary systems and 534 planets in total. Among them, the 13017 dwarfs host 285 planetary systems and 407 planets, while the 5921 subgiants host 107 planetary systems and 127 planets.
Figure~\ref{fig:hr} shows the distribution of our sample stars in the Kiel diagram, overlaid on the full LAMOST-Kepler-Gaia sample.

\section{results} \label{sec:floats}
\subsection{Stellar Age}

Figure~\ref{fig:age_distribution} illustrates the age precision and distribution of our final stellar sample. Panel~(a) shows that $\sim$73\% (13750) sample stars have age uncertainties below 20\%, and 5240 stars have uncertainties $\sim$ 20\%-50\%. The total sample has a mean age uncertainty of 15\% and a median of 10\%, indicating generally high age precision. Separating by evolutionary stage, the dwarf sample exhibits a mean age uncertainty of 18\% (median 15\%), whereas the subgiant sample achieves significantly better precision, with a mean of 8\% and a median of 6\%. Panel~(b) presents the age distribution, which peaks prominently at $\sim$4.1~Gyr, suggesting that the majority of stars in our sample are young. These results demonstrate that our stellar ages are well constrained and precise. The high quality of the age estimates provides a robust foundation for subsequent analyses of how planetary and stellar properties evolve as a function of age.

\subsection{Galactic Chemical Evolution and Planet Formation}

\begin{figure*}[htbp]
    \centering
    \includegraphics[width=\textwidth]{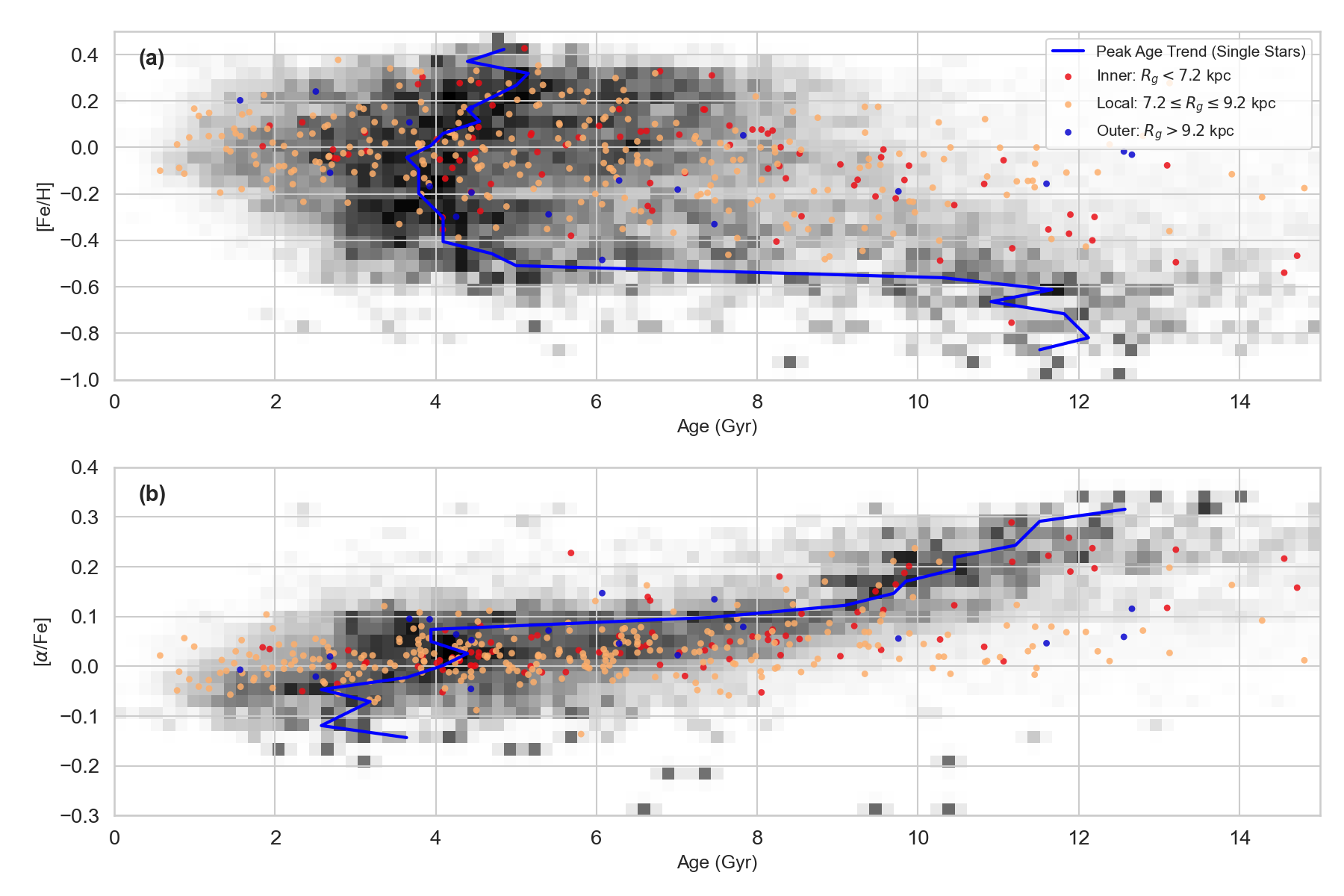}
    \caption{
    \textbf{(a)}~Normalized age distributions $p(\tau\,|\,[\mathrm{Fe/H}])$ for single stars, binned in [Fe/H] and smoothed using Gaussian filtering. Each vertical slice is normalized by its peak value. The blue line marks the peak age in each bin. Colored points show planet-hosting stars, classified by guiding-center radius $R_g$: inner ($R_g < 7.2$\,kpc), local ($7.2 \leq R_g \leq 9.2$\,kpc), and outer ($R_g > 9.2$\,kpc).  
    \textbf{(b)}~Same as (a), but for [$\alpha$/Fe] instead of [Fe/H]. 
    }
    \label{fig:age_feh}
\end{figure*}

\begin{figure*}[htbp]
    \centering
    \includegraphics[width=\textwidth]{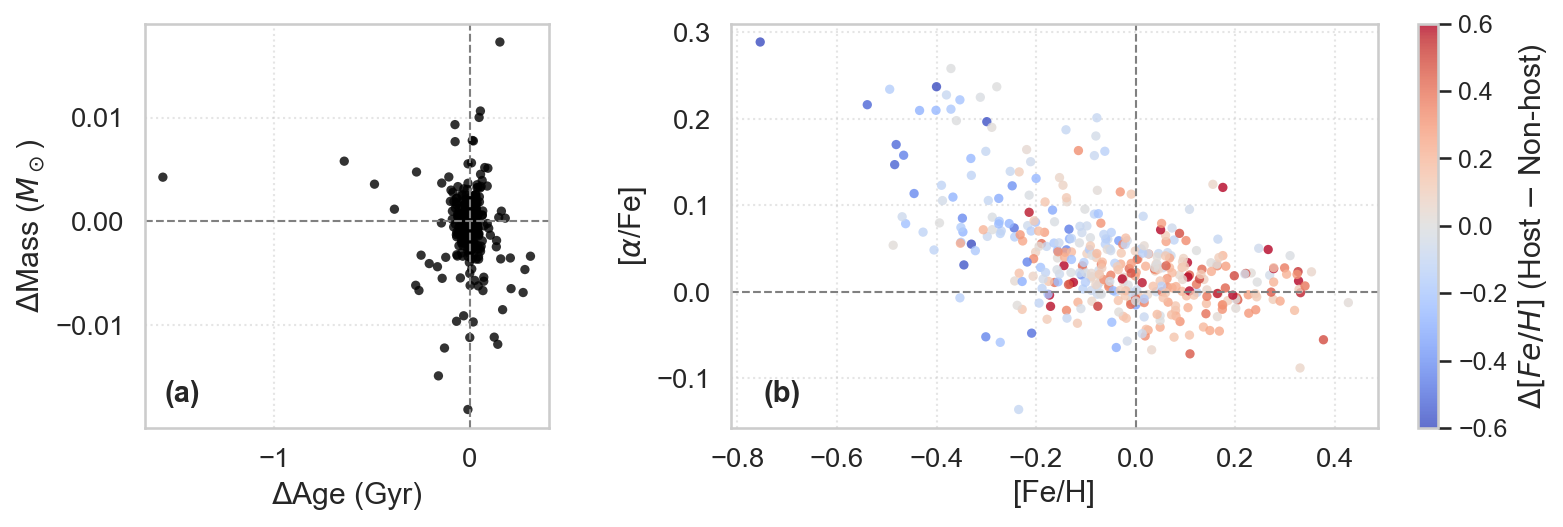}
    \caption{
    Comparison of [Fe/H] and [$\alpha$/Fe] differences between stars hosting planets and non-host stars with similar mass and age.
    (\textbf{a}) Scatter plot of differences in stellar age ($\Delta$Age) versus stellar mass ($\Delta$Mass) between host stars and their nearest non-host stars.
    (\textbf{b}) Distribution of host stars in the [Fe/H] versus [$\alpha$/Fe] plane, colored by the difference in [Fe/H] between hosts and matched non-hosts.
    }
    \label{fig:agemass}
\end{figure*}

In studies of Galactic disk stellar populations, it is common practice to separate samples into old, $\alpha$-enhanced thick disk stars and younger, low-$\alpha$ thin disk stars, based on the widely accepted view that these populations have distinct evolutionary histories. The thick disk is traditionally understood to have undergone a rapid star formation epoch characterized by a short formation timescale, whereas the thin disk experienced a more prolonged and quiescent formation period. However, recent investigations increasingly suggest that the thin disk evolution may be influenced by minor merger accretion events, resulting in a star burst in the past 4 Gyr \citep{2020NatAs...4..965R,2025NatCo..16.1581S}.

Motivated by these considerations, this work does not adopt a strict thick/thin disk classification but instead presents an integrated view of the Galactic chemical and age evolution. Panel (a) of Figure \ref{fig:age_feh} reveals a complex trend of [Fe/H] as a function of stellar age for total sample. For stars younger than 6 Gyr, a distinct ``V-shape" \citep{2018MNRAS.477.2326F,2019MNRAS.489.1742F,2022MNRAS.512.2890L,2022Natur.603..599X,2023MNRAS.523.1199S} structure emerges, with its minimum located near [Fe/H] $\approx$ –0.2. The stars in the upper branch of the ``V-shape" show increasing age with higher [Fe/H], whereas stars in the lower branch and those older than 6 Gyr exhibit the expected inverse correlation where higher [Fe/H] corresponds to younger ages. 

Previous studies \citep{2018MNRAS.477.2326F,2019MNRAS.489.1742F} have proposed that the ``V-shape" structure in the age-[Fe/H] relation arises primarily from radial migration, whereby metal-rich stars originate in the inner disk and migrate outward, while metal-poor stars formed in the outer disk migrate inward. However, alternative interpretations based on both observations and simulations suggest that late satellite infall may be responsible \citep{2021MNRAS.508.4484J,2022MNRAS.512.4697L,2022MNRAS.512.2890L,2025NatCo..16.1581S} . In this scenario, the minor merger introduces metal-poor gas into the disk, triggering the formation of a population of young stars with lower metallicity than the pre-existing disk stars. As a result, the age-[Fe/H] relation develops a secondary branch offset from the original trend, producing the observed ``V-shape" with a turning point where the two branches meet (see Figure 5 of \citet{2022MNRAS.512.4697L}).


Notably, among host stars younger than 8~Gyr, about $86\%$ lie on the upper branch of the ``V-shape" (with [Fe/H] $>-0.2$), suggesting a potential link between planet and the origins of this structure. The $R_{\mathrm{guide}}$ distribution further indicates that $\sim$76\% of host stars younger than 8 Gyr are associated with the local disk, while $\sim$19\% originate from the inner disk. Given that minor merger events primarily affect star formation in the outer disk \citep{2024MNRAS.527.4505D,2025NatCo..16.1581S}, we infer that planet-hosting stars with small $R_{\mathrm{guide}}$ are more likely shaped by radial migration rather than by merger-induced star burst. These systems likely formed in the metal-rich inner disk and were subsequently brought to the solar neighborhood through radial migration.

Panel~(b) of Figure~\ref{fig:age_feh} illustrates the relation between [$\alpha$/Fe] and stellar age, showing that nearly 87\% of host stars have relatively low $\alpha$-enhancement ([$\alpha$/Fe] $<0.1$). Notably, there is an absence of host stars with [$\alpha$/Fe] $< -0.1$, suggesting a potential role of $\alpha$-elements in planet formation. At fixed [Fe/H], an increase in [$\alpha$/Fe] results in a higher overall metallicity $Z$, which could provide the necessary conditions for planet formation. This interpretation is consistent with the abrupt lower limit of [X/H] reported by \citet{2012A&A...545A..32A}, indicating a possible overall metallicity threshold below which planet formation is suppressed. However, given the scarcity of stars with [$\alpha$/Fe] $<-0.1$ in our sample (83 stars), we cannot fully exclude the possibility that the absence of hosts in this regime arises from selection effects. Finally, we emphasize that stars without detected planets are classified as “non-hosts” in this study, although some may in fact host planets that remain undetected.

At ages $>10$~Gyr, nearly 59\% of host stars have [Fe/H] $>-0.2$. The 46 host stars in this age range have an average [Fe/H] of $-0.19$, compared to $-0.25$ for the 1878 non-host stars, indicating an overall [Fe/H] enhancement of $\sim0.06$~dex in the host stars. Furthermore, among these old host stars, $\sim$61\% belong to the local disk population (orange points) and exhibit [$\alpha$/Fe] values lower by $\sim$0.06~dex compared to non-host stars. This difference likely reflects their comparatively high [Fe/H] (see also panel~(a) of Figure~\ref{fig:age_feh}), reinforcing the view that metal-rich environments are more favorable for planet formation. In contrast, about 33\% of old hosts are associated with the inner disk (red points), where they exhibit [$\alpha$/Fe] values higher by $\sim$0.06~dex than non-host stars, indicative of enhanced $\alpha$-element abundances.

These results suggest that both old and young host stars tend to form in metal-rich environments, highlighting a stronger correlation between [Fe/H] and planet formation than with [$\alpha$/Fe], as the contribution of [$\alpha$/Fe] to the overall metallicity $Z$ is limited when [Fe/H] is fixed.

\begin{figure*}[htbp]
    \centering
    \includegraphics[width=0.95\textwidth]{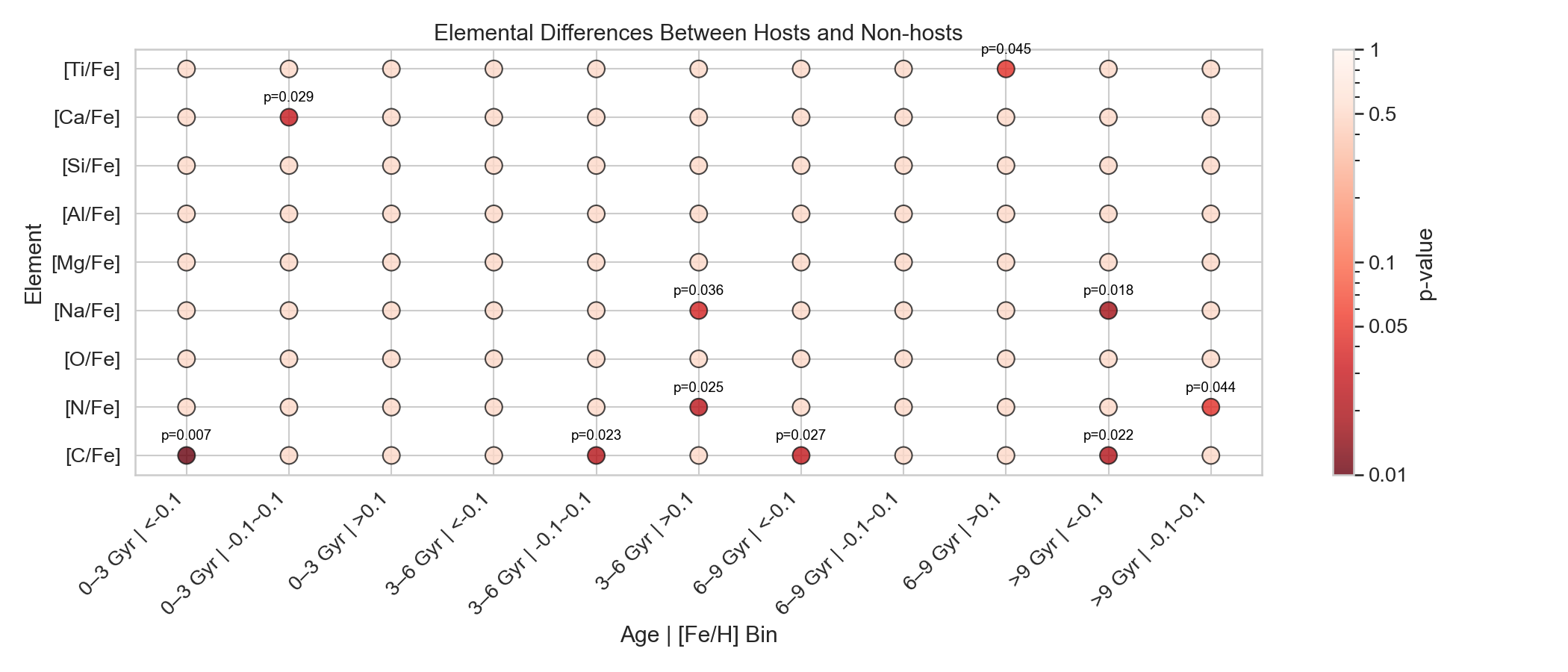}
    \caption{Elemental differences between planet-hosting and single stars across stellar age and [Fe/H] bins. 
    Each point represents the $p$-value from a two-sample Kolmogorov--Smirnov (KS) test comparing the abundance distributions of stars with and without planets. 
    Points with $p < 0.05$ indicate statistically significant differences.}
    \label{fig:ks}
\end{figure*}

\begin{figure*}[htbp]
    \centering
    \includegraphics[width=0.95\textwidth]{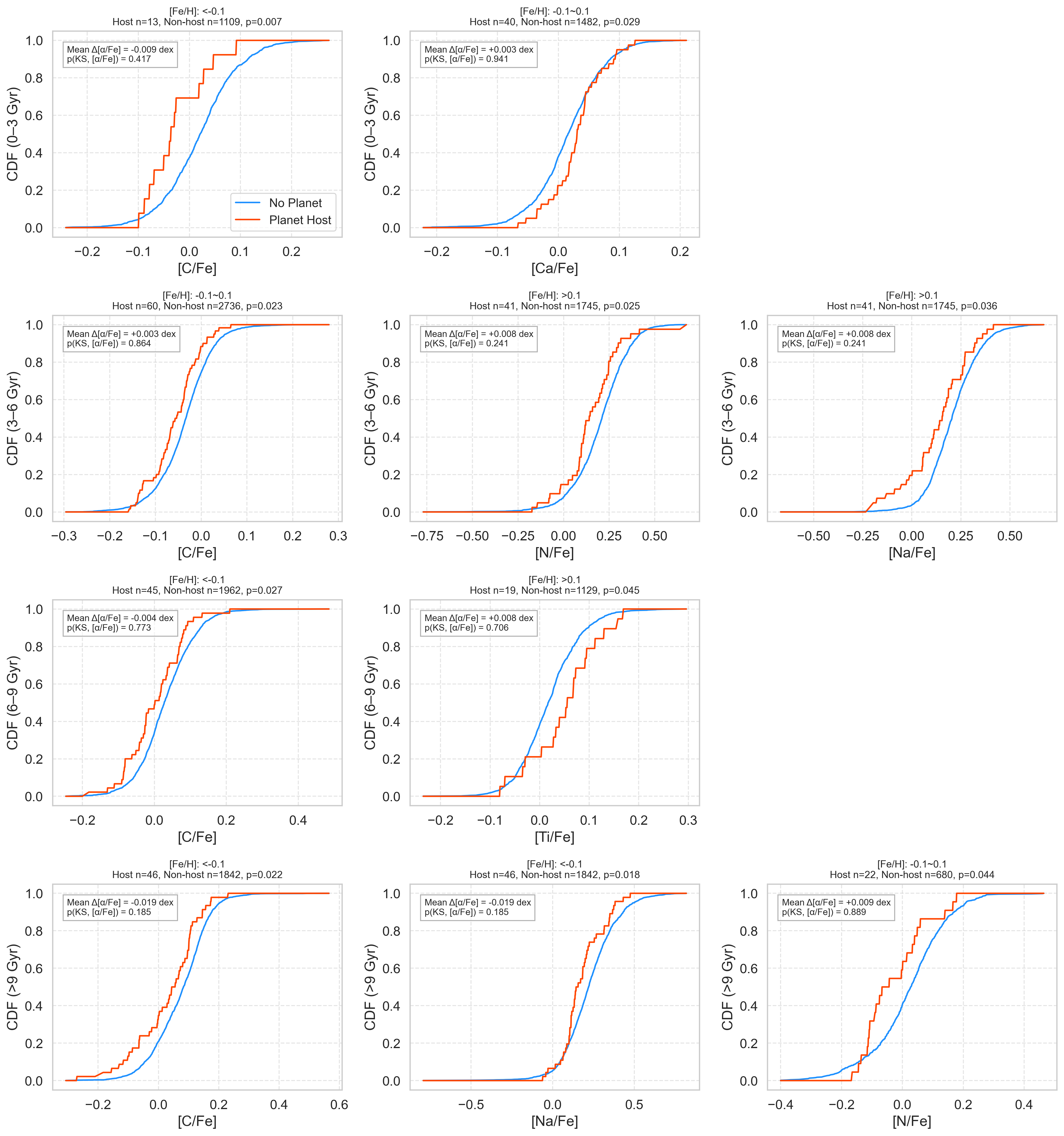}
    \caption{Cumulative distribution functions (CDFs) of elemental abundances for stars with and without planets, separated by stellar age and [Fe/H] bins. Only elements exhibiting statistically significant differences ($p < 0.05$) in a given bin are shown. Planet-hosting stars are indicated in red, and single stars in blue. Each subplot corresponds to one age bin (rows) and one significant element (columns), with [Fe/H] bin, sample sizes, and KS $p$-value labeled above each curve. In addition, the mean difference in [$\alpha$/Fe] between host and non-host stars 
($\langle \Delta[\alpha/\mathrm{Fe}] \rangle = \langle [\alpha/\mathrm{Fe}]_\mathrm{host}\rangle - \langle[\alpha/\mathrm{Fe}]_\mathrm{non-host} \rangle$) 
and the KS $p$-value for [$\alpha$/Fe] are annotated in each subplot.}
    \label{fig:cdf_elem}
\end{figure*}

\subsection{Comparison of Stellar Properties between Host and Non-host Stars}

Previous studies have shown that stars with higher [Fe/H] are more likely to host giant planets \citep{2022AJ....164...60S}, whereas the formation of terrestrial planets may not require enhanced metallicity \citep{2012Natur.486..375B}. These findings highlight the importance of [Fe/H] in planet formation, making it important to examine the [Fe/H] distribution of stars with and without planets. However, stellar mass—which correlates with the total mass of the protoplanetary disk \citep{2013ApJ...771..129A, 2016ApJ...831..125P}—and stellar age—which reflects the evolutionary stage of the system—are also key parameters. Since both mass and age are related to [Fe/H], they may introduce biases when comparing [Fe/H] distributions between host and non-host stars.

To isolate the effects of [Fe/H] from those of mass and age, we construct a control sample of non-host stars closely matched in these parameters to the 392 host stars in our sample. Specifically, for each host star, we select the non-host star with the closest stellar age and mass using the \texttt{NearestNeighbors} algorithm implemented in \texttt{scikit-learn}. We further restrict the matching pool to stars with fractional uncertainties below 20\% in age and 10\% in mass to ensure the robustness of the comparison. Although this constraint slightly reduces the proximity of some matches, it guarantees that the stellar parameters of the control sample are well constrained. As shown in Figure~\ref{fig:agemass}(a), the age and mass differences between matched pairs are typically within 0.3~Gyr and 0.005~$M_\odot$, respectively, allowing for a fair comparison of [Fe/H] and [$\alpha$/Fe] between host and non-host stars.

\begin{figure*}[htbp]
    \centering
    \includegraphics[width=0.95\textwidth]{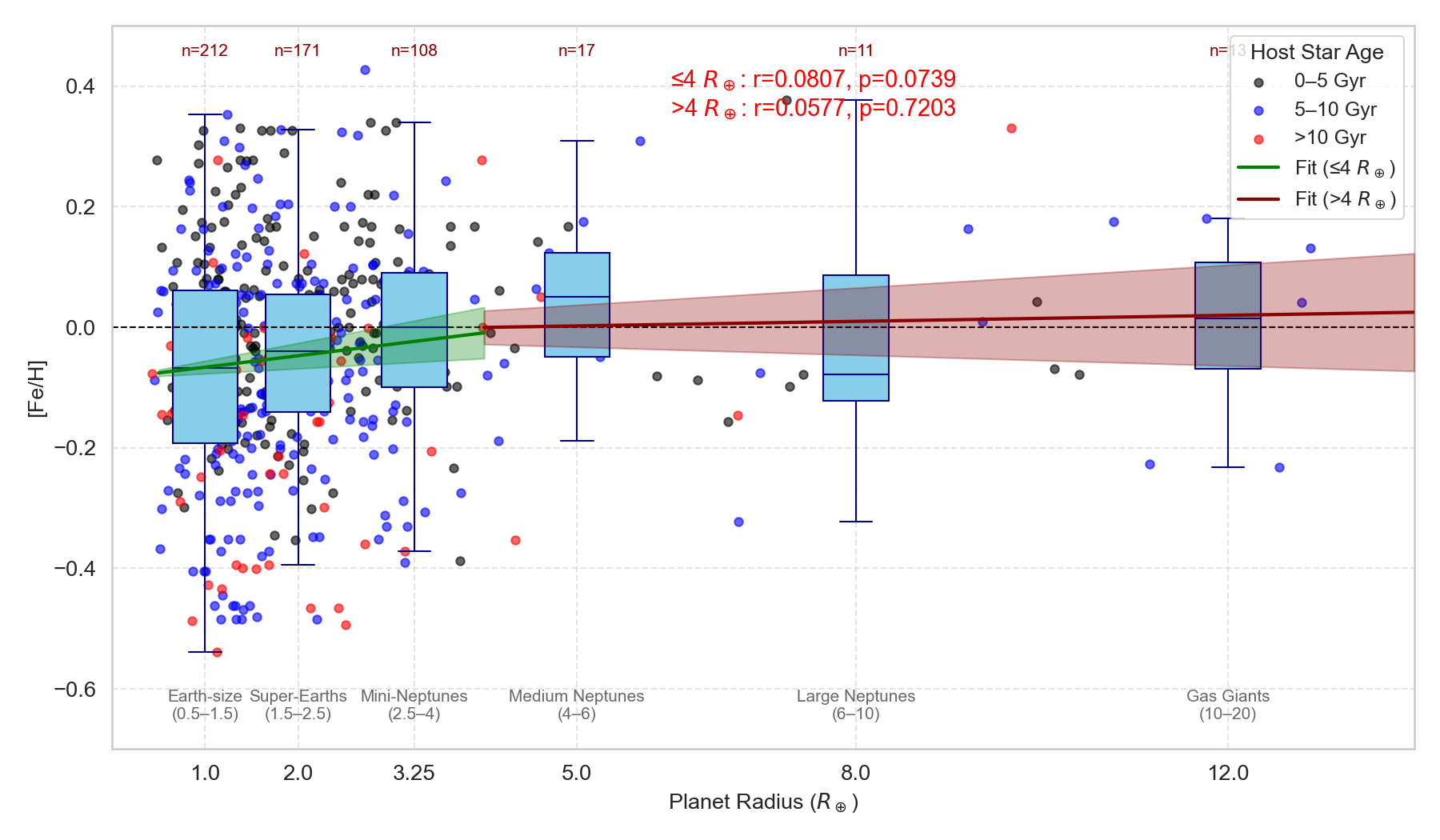}
    \caption{
        Host star metallicity [Fe/H] as a function of planet radius for different stellar age groups.
        Each dot represents an individual host star, color-coded by age: black for young stars ($\leq 5$ Gyr), blue for intermediate-age stars (5–10 Gyr), and red for old stars ($> 10$ Gyr). 
        A small horizontal jitter ($\leq 0.1$) is added to the planet radius to reduce point overlap.
        The background boxplots show the distribution of [Fe/H] in each planet radius category across all ages.
        Segmented linear regression fits are overplotted for small planets ($\leq 4~R_\oplus$, green) and large planets ($>4~R_\oplus$, dark red), with shaded areas representing 1$\sigma$ uncertainties of the fit.
        Pearson correlation coefficients ($r$) and p-values are indicated in red text.
        Sample sizes per bin are labeled above the boxplots, and planet radius classifications are noted along the bottom.
    }
    \label{fig:feh_vs_radius}
\end{figure*}

\begin{figure*}[htbp]
    \centering
    \includegraphics[width=0.95\textwidth]{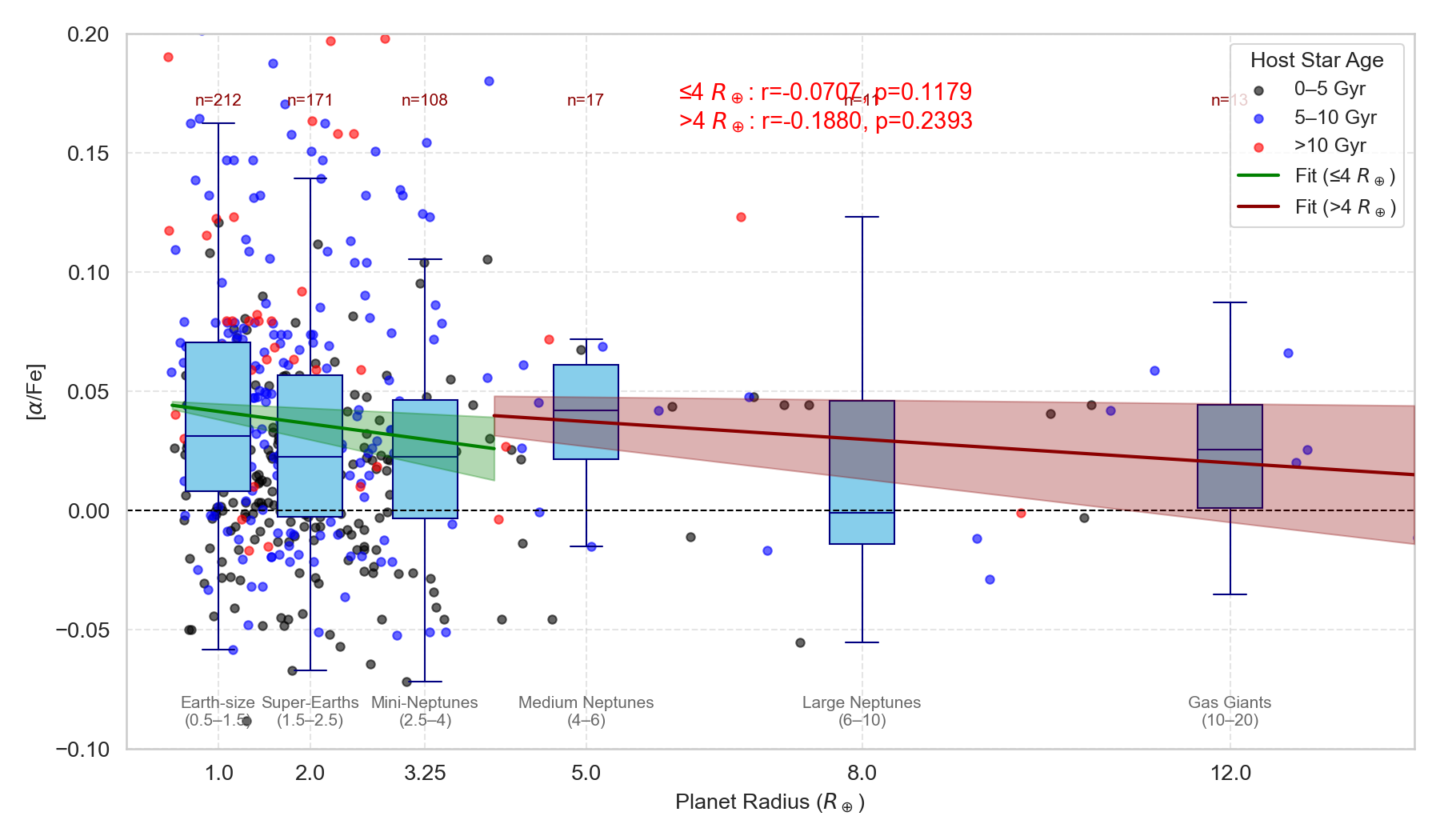}
    \caption{
        Same as Figure \ref{fig:feh_vs_radius}, but for host star [$\alpha$/Fe].
    }
    \label{fig:alpha_fe_vs_radius}
\end{figure*}

Figure~\ref{fig:agemass}(b) shows the distribution of host stars in the [$\alpha$/Fe]–[Fe/H] plane, with the colorbar indicating the [Fe/H] difference relative to matched non-host stars ($\Delta$[Fe/H] = [Fe/H]$_\mathrm{host}$ – [Fe/H]$_\mathrm{non-host}$). In the regime $\mathrm{[Fe/H]} > -0.2$, about 67\% of host stars have higher [Fe/H] than non-host stars, with an average enhancement of $\sim$0.13\,dex. Conversely, for $\mathrm{[Fe/H]} < -0.2$, nearly 77\% of host stars show lower [Fe/H], with an average deficit of $\sim$0.16\,dex relative to non-host stars. Although planet formation appears to be favored in metal-rich environments as mentioned above, [Fe/H] is not a key factor for planet formation in metal-poor conditions. This also supports the idea that high [Fe/H] is not a necessary condition for the formation of small planets \citep{2012Natur.486..375B,2023AJ....166...91S}.

To further explore the chemical environment of planet-hosting stars, we analyze the distributions of individual elemental abundances within bins defined by both stellar age and [Fe/H]. Stellar age is divided into four groups: 0–3, 3–6, 6–9, and $>9$ Gyr, while [Fe/H] is categorized into three ranges: $<-0.1$ (metal-poor), $-0.1 \leq \rm[Fe/H] \leq 0.1$ (solar-like), and $>0.1$ (metal-rich). We focus on elements with different condensation temperatures that are important for planet formation: volatile elements (C, N, O), moderately volatile (Na), and refractory elements (Mg, Al, Si, Ca, Ti). These elements play distinct roles in shaping planetary composition. For instance, the formation of rocky planets is strongly linked to C, O, Mg, and Si, while the formation of gas and ice giants depends more on volatile species such as C, N, O, and S \citep{2023MNRAS.524.6295P}. Moreover, the stellar C/O ratio is a critical parameter: a higher C/O ratio favors the formation of carbonates, whereas for lower C/O ratios, the Mg/Si ratio regulates the dominant silicate mineralogy in planetary interiors \citep{2016ApJ...831...20B}.

Figure~\ref{fig:ks} presents the results of two-sample Kolmogorov--Smirnov (KS) tests comparing elemental abundances between planet-hosting and single stars across different age and [Fe/H] bins. We restrict our analysis to bins with more than 10 planet-hosting stars to ensure statistical robustness. Figure~\ref{fig:cdf_elem} presents the cumulative distribution functions (CDFs) of elemental abundances for stars with and without planets. Only elements showing statistically significant differences ($p < 0.05$) in each bin are included. We note that even within the similar age and [Fe/H] bins, [$\alpha$/Fe] may vary due to the Galactic birth environment and chemo-dynamical evolution \citep{2013A&A...558A...9M,2014A&A...562A..71B}, and such variation can influence the volatile and refractory abundance contrasts. To assess the potential impact of such effects, we examine the [$\alpha$/Fe] distributions in each bin. We find that for all age and [Fe/H] groups, the [$\alpha$/Fe] distributions of host and non-host stars show no statistically significant differences, with mean differences smaller than 0.02~dex. This indicates that the impact of [$\alpha$/Fe] on the abundance contrasts is limited.

For volatile elements such as [C/Fe], [N/Fe], and [Na/Fe], planet-hosting stars generally exhibit systematically lower abundances compared to single stars in certain age and [Fe/H] bins. Specifically, [C/Fe] shows significant depletion in planet hosts in four bins: 0--3~Gyr with [Fe/H] $< -0.1$, 3--6~Gyr with $-0.1 \leq$ [Fe/H] $\leq 0.1$, 6--9~Gyr with [Fe/H] $< -0.1$, and $>9$~Gyr with $-0.1 \leq$ [Fe/H] $\leq 0.1$. [N/Fe] is significantly lower in hosts in the 3--6~Gyr bin with [Fe/H] $> 0.1$ and in the $>9$~Gyr bin with $-0.1 \leq$ [Fe/H] $\leq 0.1$, while [Na/Fe] is depleted in the 3--6~Gyr bin with [Fe/H] $> 0.1$ and in the $>9$~Gyr bin with [Fe/H] $< -0.1$. In contrast, for refractory elements such as [Ca/Fe] and [Ti/Fe], we observe that planet-hosting stars can exhibit higher abundances in certain bins: [Ca/Fe] is significantly enhanced in hosts in the 0--3~Gyr bin with $-0.1 \leq$ [Fe/H] $\leq 0.1$, and [Ti/Fe] shows higher values in the 6--9~Gyr bin with [Fe/H] $< -0.1$. These results suggest that volatile elements are generally depleted in planet-hosting stars, whereas some refractory elements can be enriched in specific age and metallicity ranges.

Our results indicate that planet-hosting stars exhibit the most pronounced C depletion among volatile elements, which is observed across the full range of stellar ages. N and Na, in contrast, show depletion only in older stars with higher [Fe/H]. For refractory elements, enhancements are seen only in a few specific bins for [Ca/Fe] and [Ti/Fe], suggesting that most refractory elements remain relatively stable. These findings are consistent with \citet{2014A&A...562A..27T}, who proposed that volatile elements are more sensitive to protoplanetary disk conditions than refractories. As noted by \citet{2011ApJ...743L..16O}, regions between the H$_2$O and CO snowlines can regulate the accretion and migration of volatile-rich material, thereby altering observed surface abundances while leaving refractory elements largely unaffected. Additionally, planetary ingestion \citep{2023MNRAS.521.2969B,2024Natur.627..501L,2025MNRAS.538.2408Y} may also contribute to the observed enhancement of refractory elements. 

Our results are also inconsistent with some previous work. For example, \citet{2022AJ....164..181U} found comparable [C/Fe]--[Fe/H] trends for planet-hosting and non-host stars, and \citet{2017A&A...599A..96S} suggested that planet hosts tend to be more carbon-rich. Many previous studies report that, compared with 80--90 per cent of solar twins, the Sun exhibits refractory depletion relative to iron and more volatile elements such as carbon and oxygen \citep[e.g.,][]{2009ApJ...704L..66M, 2009A&A...508L..17R, 2020MNRAS.493.5079B,2024ApJ...965..176R}. The discrepancies between our findings and previous studies may arise from a combination of factors, including sample selection (e.g., solar twins versus broader FGK populations), abundance analysis methods (high-resolution homogeneous measurements versus catalog-based values), and the nature of planets hosted by the stars. These methodological and observational differences highlight that abundance signatures of planet formation are sensitive to both stellar context and measurement precision.

\subsection{Stellar Properties and Planet Size}

A key question in the study of star--planet correlations is how the properties of host stars influence the radii of their planets---that is, what kinds of stars are more likely to host larger planets. Figure~\ref{fig:feh_vs_radius} presents the distribution of host star [Fe/H] as a function of planet radius, separated by stellar age groups. The sample is dominated by planets in the Earth-size to mini-Neptune regime ($\leq 4~R_\oplus$). The segmented linear regression analysis shows that for small planets ($\leq 4~R_\oplus$), the correlation between [Fe/H] and planet radius is weak and statistically insignificant ($r = 0.081$, $p = 0.0739$). For larger planets ($> 4~R_\oplus$), the correlation is even weaker ($r = 0.058$, $p = 0.7203$). These results indicate that, although [Fe/H] is important for planet formation, it does not show a significant correlation with planet size in our sample. This is broadly consistent with previous studies showing little to no metallicity correlation for sub-Neptunes and rocky planets \citep[e.g.,][]{2018AJ....155...89P}. Previous studies have established that the occurrence rate of giant planets increases with host star metallicity \citep[e.g.,][]{2004A&A...415.1153S,2005ApJ...622.1102F,2010PASP..122..905J}, with the positive metallicity correlation being most pronounced for Jovian-mass planets \citep[e.g.,][]{2005ApJ...622.1102F}. In our sample, we do not find a significant correlation between host star [Fe/H] and planet radius for giant planets. This result can be attributed to two main factors: first, the number of detected large planets is very small ($\geq 4~R_\oplus$, $n=41$; $\geq 10~R_\oplus$, $n=13$); second, our analysis is restricted to the detected planets, which differs from studies focusing on overall planet occurrence. In our sample, the mean [Fe/H] of host stars for small planets ($\leq 4~R_\oplus$, $n=491$) is $-0.051$~dex, whereas for large planets ($\geq 4~R_\oplus$, $n=41$) it is 0.010~dex. Despite the limited number of larger planets, these results qualitatively suggest that they preferentially orbit more metal-rich stars. Notably, none of the host stars of planets larger than $4~R_\oplus$ have [Fe/H] below $-0.4$~dex, consistent with the findings of \citet{2012Natur.486..375B}.

We also examined host star [$\alpha$/Fe] as a function of planet radius (Figure~\ref{fig:alpha_fe_vs_radius}). Again, no significant correlation is found: for small planets ($\leq 4~R_\oplus$), $r = -0.071$, $p = 0.1179$, and for large planets ($> 4~R_\oplus$), $r = -0.188$, $p = 0.2393$. While [$\alpha$/Fe] can contribute to overall metallicity, these weak correlations suggest that $\alpha$-enhancement does not play a important role in determining planet size. However, we note that host stars with [$\alpha$/Fe] $> 0.1$ predominantly harbor small planets, suggesting that $\alpha$-enhanced environments are conducive to their formation; while host stars of large planets (radius $> 5~R_\oplus$) tend to have low [$\alpha$/Fe]. These findings suggest that the formation of giant planets may proceed through mechanisms that are largely independent of [Fe/H] and [$\alpha$/Fe].

\section{Summary} \label{sec:summary}

In this study, we determine precise stellar ages for 18890 dwarfs and subgiants from the LAMOST–Kepler–Gaia sample, with an mean relative uncertainty around 15\% and a median of 10\%, peaking at 4.1~Gyr. Within the framework of Galactic chemical evolution, we find that nearly 86\% of planet-hosting stars occupy the metal-rich branch ([Fe/H] $>$ $-0.2$) of the characteristic ``V-shape" structure in the age-metallicity relation of the Galactic disk. These stars are primarily associated with the local and inner disk populations. Their distribution indicates that about 19\% of hosts younger than 8 Gyr may have undergone radial migration, carrying their planetary systems from the metal-rich inner disk to the solar neighborhood. Furthermore, among stars older than 10 Gyr, host stars are on average more metal-rich than non-host stars, with nearly 59\% having $\mathrm{[Fe/H]} > -0.2$, suggesting that metal-rich environments are conducive to planet formation.

By comparing host stars with non-host stars of similar age and mass, we find that in the high-metallicity regime ([Fe/H] $>-0.2$), about 67\% of host stars are more metal-rich than non-hosts, with an average enhancement of $\sim$0.13\,dex. Conversely, in the low-metallicity regime ([Fe/H] $<-0.2$), nearly 77\% of host stars are more metal-poor, exhibiting an average deficit of $\sim$0.16\,dex relative to non-hosts. These results indicate that while planet formation is generally favored in metal-rich environments, [Fe/H] does not appear to be a dominant factor in planet formation under metal-poor conditions, supporting the idea that high metallicity is not a necessary requirement for the formation of small planets.

We divided the sample into age and [Fe/H] bins and compared elemental abundances between host and non-host stars. Our analysis shows that [C/Fe] is consistently depleted in planet-hosting stars across all ages, indicating a systematic carbon deficiency. Other volatile elements, such as N and Na, exhibit depletion only in specific age and metallicity bins. In contrast, refractory elements like Ca and Ti are enhanced only in a few bins, with most refractories remaining relatively stable. These results support the idea that volatile elements are more sensitive to planet formation processes, while some refractory elements may experience enhancement under certain conditions.

Host star [Fe/H] and [$\alpha$/Fe] show only weak correlations with planet radius. Stars with [$\alpha$/Fe] $>0.1$ predominantly host small planets ($\leq 4~R_\oplus$), suggesting that $\alpha$-enhanced environments may favor the formation of small planets. For larger planets ($\geq 4~R_\oplus$), although the number of detected large planets is limited, the mean [Fe/H] of their host stars is slightly higher (0.010~dex) than that of smaller planets ($-0.051$~dex), indicating that larger planets tend to orbit more metal-rich stars.

\begin{acknowledgements}
This work is based on data acquired through the Guoshoujing Telescope. Guoshoujing Telescope (the Large Sky Area Multi-Object Fiber Spectroscopic Telescope; LAMOST) is a National Major Scientific Project built by the Chinese Academy of Sciences. Funding for the project has been provided by the National Development and Reform Commission. LAMOST is operated and managed by the National Astronomical Observatories, Chinese Academy of Sciences. This work has made use of data from the European Space Agency (ESA) mission Gaia (\url{https://www.cosmos.esa.int/gaia}), processed by the Gaia Data Processing and Analysis Consortium (DPAC, \url{https://www.cosmos.esa.int/web/gaia/dpac/consortium}). Funding for the DPAC has been provided by national institutions, in particular the institutions participating in the Gaia Multilateral Agreement. We acknowledge the entire Kepler team and everyone involved in the Kepler mission. Funding for the Kepler mission is provided by NASA's Science Mission Directorate.
This work was supported by the National Natural Science Foundation of China (Grants No.12403037).
\end{acknowledgements}

\appendix
\section{Comparison of stellar ages} \label{app:age_comparison}

\begin{figure*}[htbp]
    \centering
    \includegraphics[width=0.95\textwidth]{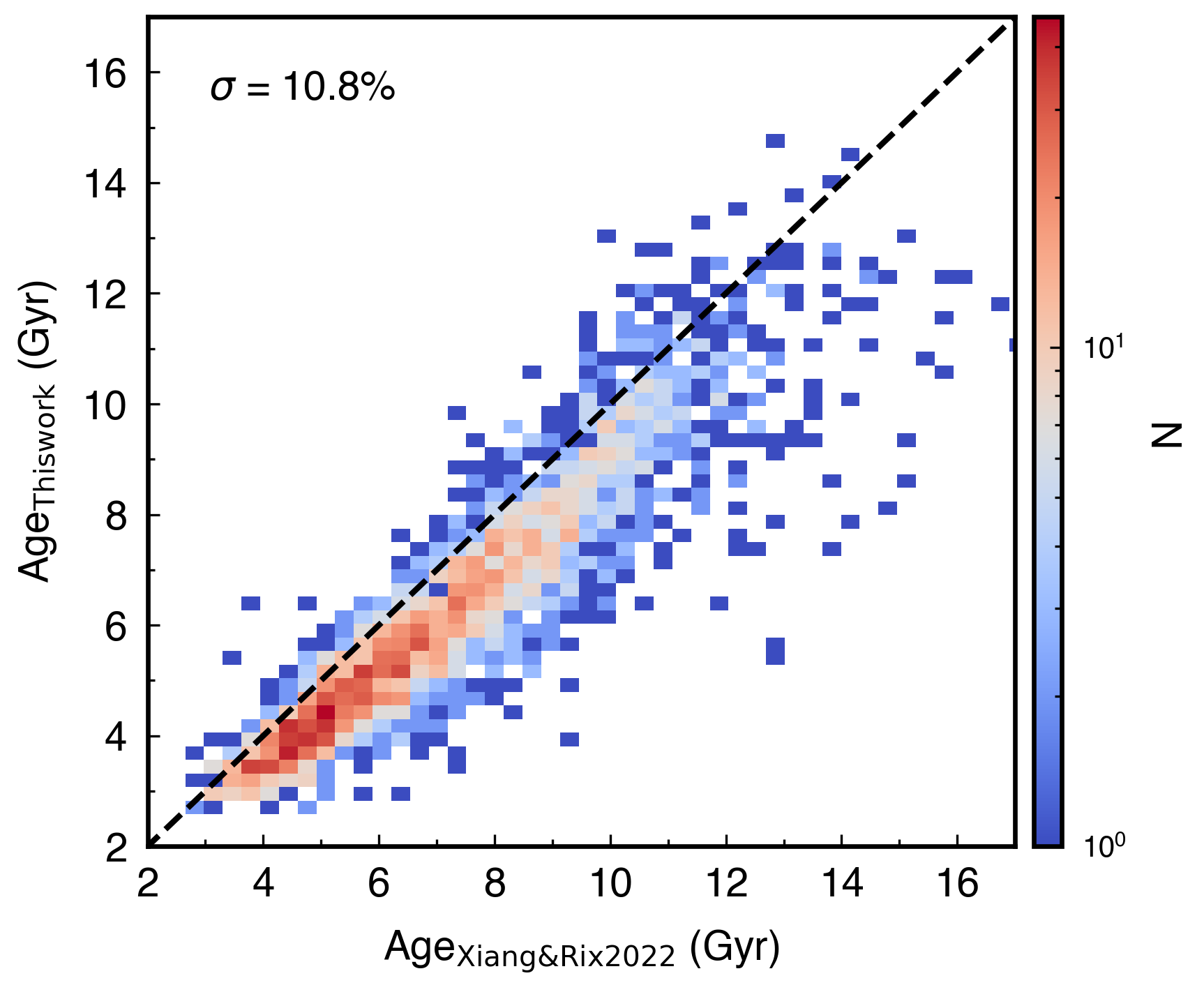}
    \caption{Comparison of ages for the 2866 stars in the cross-matched sample between our dataset and the subgiant sample from from \cite{2022Natur.603..599X}. The black line indicates the 1:1 relation. The mean age offset is $-$12.9\%, with a scatter of 10.8\%.}
    \label{fig:age_comparison}
\end{figure*}

Figure \ref{fig:age_comparison} compares the age estimates of 2866 stars in the cross-matched sample between our dataset and the subgiant sample from \cite{2022Natur.603..599X}. In \cite{2022Natur.603..599X}, stellar ages were derived by fitting Gaia parallaxes, LAMOST spectroscopic parameters ($T_{\rm eff}$, $M_{\rm K}$, [Fe/H], and [$\alpha$/Fe], obtained from LAMOST DR7 spectra via the data-driven Payne approach), and Gaia and 2MASS photometry to YY isochrones \citep{2004ApJS..155..667D} using a Bayesian framework. As shown in Figure \ref{fig:age_comparison}, our age estimates are systematically younger than those of \cite{2022Natur.603..599X} by $\sim$12.9\%, with a scatter of 10.8\%. This offset likely reflects several methodological differences: the adopted $T_{\rm eff}$ and [Fe/H] values, the use of $\alpha$-enhanced models from \citet{2023MNRAS.523.1199S} rather than the YY isochrones, and differing extinction treatments-our analysis employs the extinction map of \citet{2019ApJ...887...93G}, while \citet{2022Natur.603..599X} derived their own values. Collectively, these differences account for the systematic shift between the two sets of age determinations.


\bibliography{sample701}{}

\bibliographystyle{aasjournalv7}

\end{document}